\documentclass[journal]{IEEEtran}

\usepackage[latin1]{inputenc}
\usepackage{setspace}
\usepackage{graphicx, amssymb, amsmath, amsfonts, amsthm}
\usepackage[usenames,dvipsnames]{color}
\usepackage{stfloats}
\usepackage{subcaption}
\usepackage{comment}
\usepackage{multicol}

\allowdisplaybreaks
\IEEEoverridecommandlockouts

\begin{document}
\title{A Novel Temperature-based Model for SWIPT}

\author{Elio Faddoul, \IEEEmembership{Graduate Student Member, IEEE}, Yuan Guo, \IEEEmembership{Graduate Student Member, IEEE},\\ Christodoulos Skouroumounis, \IEEEmembership{Senior Member, IEEE}, and Ioannis Krikidis, \IEEEmembership{Fellow, IEEE}\vspace*{-4mm}
\thanks{This work received funding from the European Research Council (ERC) under the European Union's Horizon 2020 research and innovation programme (Grant agreement No. 819819) and the European Union's Horizon Europe programme (ERC, Grant agreement No. 101112697).}\thanks{The authors are with the IRIDA Research Centre for Communication
Technologies, Department of Electrical and Computer Engineering, University of Cyprus, Cyprus (email: \{efaddo01, yguo0001, cskour03, krikidis\}@ucy.ac.cy).}}


\maketitle

\begin{abstract}
In this letter, a novel communication paradigm for simultaneous wireless information and power transfer (SWIPT) is proposed, which leverages the thermal characteristics of electromagnetic signals. In particular, the proposed scheme exploits the inherent thermal dynamics of electromagnetic signals, enabling the seamless integration of information decoding and energy harvesting (EH). As a consequence, in contrast to conventional SWIPT techniques, the proposed model eliminates the need to divide the received signal into orthogonal components. By exploiting the thermal correlation between consecutive time slots, the communication channel is converted to a virtual multiple-input multiple-output (MIMO) channel with memory. We evaluate the achievable rate of the proposed temperature-modulated channel for uniform and exponential input distributions and assess its performance in terms of harvested energy through a non-linear harvesting model. Our numerical results reveal that the exponential distribution outperforms the uniform distribution in rate and harvested energy at low input power levels, while the uniform distribution achieves a better EH performance at high input power levels.
\end{abstract}


\section{Introduction}
\IEEEPARstart{T}HE advent of sixth-generation (6G) communication networks, characterized by extensive low-power device connectivity, raises concerns regarding the energy sustainability of these devices \cite{OZ}. Traditional solutions like battery replacement and natural resource-based energy harvesting (EH) are rendered unstable, costly, and infeasible in specific scenarios. Within the framework of the Internet of Things (IoT) and wireless sensor networks, the concept of wireless power transfer (WPT) has emerged as a promising solution to address the critical challenge of delivering power wirelessly to numerous low-power IoT devices \cite{LU}. In particular, radio-frequency (RF) signals are considered a viable energy source for WPT, demonstrating the capability to energize devices effectively even over significant distances.

Notably, RF signals have been widely employed for wireless data transmission. Consequently, the concept of simultaneous wireless information and power transfer (SWIPT), where RF signals concurrently transmit both data and energy, emerges as an appealing solution for networks featuring low-power devices \cite{KRI}. Generally, the research community has mostly focused on devices that harvest energy through a rectifying antenna, \textit{i.e.}, a diode-based circuit that converts RF signals to direct-current voltage \cite{PE}. In the context of SWIPT systems, the joint extraction of information and energy is realized by separating the information decoding and EH operations in space, time, or power, revealing the fundamental tradeoff between information and energy transfer \cite{WEI}. Hence, conventional SWIPT techniques are deemed inefficient, as they sacrifice overall system performance in the pursuit of simultaneously decoding information and harvesting energy.

In addition to the concept of SWIPT, there has been a recent surge of interest in thermal- or temperature-based communication, considering both transmission and reception aspects \cite{ZA, OZ2, TO}. Initially employed for covert (hidden) communications scenarios, the thermal channel capacity for such applications was empirically explored in \cite{ZA}. In addition, optimal transmit power policies, subject to temperature dynamics and EH constraints, were investigated over an additive white Gaussian noise channel in \cite{OZ2}. More recently, a theoretical foundation for utilizing heat conduction as a mode of communication has been established in \cite{TO}. This work emphasizes the fundamental distinctions of the thermal channel from conventional wireless communication channels, with potential advantages in nano-scale communications in order to achieve simultaneous information and power transfer.

In this paper, a novel SWIPT model is proposed that enables simultaneous information transmission and EH without splitting the available resources at the receiver, leading to an enhanced overall network performance. Indeed, the proposed scheme exploits the thermal effects induced by electromagnetic signals as a means to represent a true SWIPT system. The receiver, which is equipped with a temperature thermometer and a rectifying antenna, capitalizes on the temperature fluctuations in the RF signal for information decoding and utilizes the RF signal itself for EH. In particular, a virtual multiple-input multiple-output (MIMO) setup is constructed by considering consecutive time instances and exploiting the associated channel memory. As the temperature is always positive, our established mathematical framework effectively transforms this virtual MIMO channel into an equivalent MIMO intensity channel \cite{CHA1}. This transformation allows the utilization of established expressions to evaluate the channel capacity. To this end, we first provide an achievable rate for the proposed temperature-modulated channel by considering two different input distributions, \textit{i.e.}, the uniform and exponential distributions. Then, we assess the performance of the temperature-modulated channel in terms of harvested energy by employing a non-linear harvesting model. Numerical results indicate that, at low input power levels, utilizing an exponential input distribution results in superior rate and EH performance compared to a uniform input distribution, while at high input power levels, the uniform input distribution exhibits better EH performance. To the best of our knowledge, this is the first work that exploits the temperature dynamics of RF signals in the context of SWIPT systems.

\textit{Notation:} Lower and upper case boldface letters denote vectors and matrices, respectively; $[\cdot]^{-1}$ and $[\cdot]^{T}$ are the inverse and transpose, respectively; det$(\cdot)$ is the determinant operator; $|\cdot|$ and $\|\cdot\|$ denote the norm of a scalar and vector, respectively; $\mathbb{E}\{\cdot\}$ represents the expectation operator; $Q(\cdot)$ is the Gaussian $Q$-function and $K_{n}(\cdot)$ is the modified Bessel function of the $n$-th order and the second kind.

\section{System model}
\begin{figure}
\includegraphics[width=0.485\textwidth]{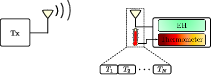}
\centering
\caption{System model of the considered temperature-modulated channel for SWIPT.}\label{fig:1}
\end{figure}

We consider a point-to-point temperature-based SWIPT wireless link consisting of a transmitter and receiver pair, both equipped with single antennas as depicted in Fig. \ref{fig:1}. At the $i$-th channel use, the source transmits the symbol $x_i$ with transmit power $S_i=|x_i|^2 $. The wireless channel exhibits Rayleigh flat-fading where $g_i$ denotes the associated channel coefficient. The receiver consists of a conventional EH circuit (rectenna) that converts the total received RF signal to DC. Furthermore, we consider that the receiver holds a local thermometer that detects the inherent temperature fluctuations due to the incident RF power. The thermometer is assumed to be co-located with the receiver's antenna, forming a thermocouple sensor that measures temperature and outputs a voltage reading \cite{TO}. Regarding the EH process, let $\mathcal{Q}(\cdot)$ denote the EH function. For the $i$-th channel use, the received power is $P_i=h_i S_i$, where $h_i \triangleq |g_i|^2$. Therefore, the associated harvested power is equal to $\mathcal{Q}(P_i)$. Further details on the information decoding process are given in the next subsection (see Section II-A).

\begin{figure*}[b]
\vspace{2mm}
\hrulefill
\vspace{2mm}
\begin{align}\tag{4} \label{mod4}
\left[\begin{array}{ccccc} T_2 \\[1mm] T_3 \\[1mm] T_4 \\[1mm] \vdots \\[1mm] T_{N+1} \end{array}\right] \!\! = \!\! \left[\begin{array}{ccccc} T_e \\[1mm] T_e \\[1mm] T_e \\[1mm] \vdots \\[1mm] T_e \end{array}\right] \!\!+\!\! \underbrace{\left[\begin{array}{ccccc}
\!\! \alpha h_1 & 0 & 0 & \cdots & 0 \!\! \\[1mm]
\!\! \alpha (1-\beta) h_1 & \alpha h_2 & 0 & \cdots & 0 \!\! \\[1mm]
\!\! \alpha (1-\beta)^2 h_1 & \alpha (1-\beta) h_2 & \alpha h_3& \cdots & 0 \!\! \\[1mm]
\!\! \vdots & \vdots & \vdots & \ddots & \vdots \!\! \\[1mm]
\!\! \alpha (1-\beta)^{N-1} h_1 & \alpha (1-\beta)^{N-2} h_2 & \alpha (1-\beta)^{N-3} h_3 & \cdots & \alpha h_N \!\!
\end{array}\right]}_{\bf{A}} \! \left[\begin{array}{ccccc} S_1 \\[1mm] S_2 \\[1mm] S_3 \\[1mm] \vdots \\[1mm] S_{N} \end{array}\right] \!\!+\!\! \left[\begin{array}{ccccc} w_1 \\[1mm] w_2 \\[1mm] w_3 \\[1mm] \vdots \\[1mm] w_{N} \end{array}\right]
\end{align}
\end{figure*}

\subsection{Thermal Dynamics at the Receiver}
In contrast to conventional SWIPT architectures that split the received signal (in the RF or baseband domain) to decode information, the proposed receiver exploits the thermal effects of the received RF signal to convey information, while using the EH circuit to extract energy from the same signal. In particular, the temperature model can be characterized as a first-order heat circuit that is driven by the electromagnetic radiation emitted due to the transmit power. Thus, the temperature dynamics at the receiver side are modeled as \cite{OZ2}
\begin{align}
\frac{dT}{dt}=\alpha P(t)-\beta \left(T(t)-T_e\right),
\label{mod}
\end{align}
where $P(t)$ is the received power at the time instant $t$, $T(t)$ is the temperature at the receiver at time instant $t$, $T_e$ is the temperature of the environment; $\alpha$ and $\beta$ are non-negative constants. By discretizing the temperature model and by assuming a normalized channel use duration (unitary duration), we can rewrite \eqref{mod} as
\begin{align}\tag{2}  \label{mod2}
T_{i+1}-T_{i}=\alpha P_{i}-\beta\left(T_{i}-T_e\right),
\end{align}
where $T_i$, $T_{i+1}$ denote the receiver's temperature at the beginning and the end of the channel use $i$, respectively, due to the reception of the RF power $P_i$;  $i=1,2,\ldots N$ and $N$ is the total number of channel uses. By assuming $T_1=T_e$  without loss of generality, the discrete model in \eqref{mod2} is simplified to
\begin{align}\tag{3} \label{mod3}
T_{i+1}=T_e+\alpha \sum_{k=1}^i (1-\beta)^{i-k}P_k + w_k,
\end{align}
where $w_k$ is the additive white Gaussian noise (AWGN) component with zero mean and variance $\sigma^2$. It is assumed that the AWGN component is independent of the input signal \cite{TO}.

\subsection{Equivalent Temperature Channel}
The proposed model for SWIPT exploits the temporally correlated thermal dynamics of the RF signal. By leveraging the thermal correlation between consecutive time slots, and by using the temperature fluctuations as a means to convey information, the communication channel is converted to a virtual MIMO channel with memory. In particular, the receiver focuses on $N$ consecutive channel instances, forming an $N \times N$ virtual MIMO channel with memory. Based on the expression in \eqref{mod3}, we can express the MIMO temperature channel as given in \eqref{mod4} at the bottom of this page. This approach captures and leverages the inherent memory effects within the channel, thereby transforming the single-input single-output channel into a virtual MIMO configuration. The uniqueness of this expression lies in the lower triangular structure of the temperature channel matrix $\bf{A}$ in \eqref{mod4}, simplifying the receiver post-coding process when assessing the achievable rate.

\subsection{Energy Harvesting}
Regarding the EH process, we consider a piecewise linear EH model where the harvested energy at the channel use $i$ is given by
\begin{equation}\tag{5}\label{EH_mod}
\mathcal{Q}(P_i)=\begin{cases}
0,& \text{if}\ {}\ {}\ {} 0\le P_i\le P_{\rm th},\\[2mm]
\eta P_i,&\text{if}\ {}\ {} P_{\rm th}< P_i \le P_{\rm sat},\\[2mm]
\eta P_{\rm sat},&  \text{if}\ {} \ {} P_i>P_{\rm sat},
\end{cases}
\end{equation}
where $0\leq \eta \leq 1$ is the RF-to-DC energy conversion efficiency, $P_{\rm th}$ and $P_{\rm sat}$ represent the sensitive threshold and the saturation power of the energy harvester, respectively.

\section{Temperature-Modulated SWIPT Model}
In this section, the characterization of the considered temperature-modulated channel for SWIPT is examined in terms of the ergodic achievable rate as well as the average harvested energy.

\subsection{Achievable Rate}
As the temperature is invariably positive, we observe that our proposed temperature-based virtual MIMO channel closely aligns with the intensity MIMO channel, which uses the signal intensity to transmit information \cite{CHA1}. Although this connection facilitates the derivation of the channel capacity, an exact determination of the capacity of such channels remains unknown, necessitating the evaluation of achievable rates. Therefore, we consider the computation of the ergodic achievable rate under the channel inversion method for two well-known input distributions, namely the uniform and the exponential distributions. It is important to mention here that the choice of the input distributions is not arbitrary. In particular, the achievable rate by employing an exponential input distribution asymptotically coincides with the upper bound, as discussed in \cite{CHA}, while the scenario with a uniform input distribution acts as a benchmark for the performance evaluation. Accordingly, the achievable rate for the temperature-modulated channel using channel inversion is given by
\begin{equation}\tag{6} \label{mod5}
R_{e}^{\rm CI}(\mathbf{A})=\sum_{i=1}^{N}\frac{1}{2N}\log_2\left(1+\frac{e\mathcal{E}_i^2}{2\pi\| \mathbf{b}_i\|^2}\right),
\end{equation}
for the exponential input distribution satisfying $\mathbb{E}\left\{S_i\right\} \leq \mathcal{E}_i$, where $\mathbf{b}_{i}^{T}$ is defined as the $i$-th row of $\mathbf{B}\triangleq \mathbf{A}^{-1}$. Note that the channel inversion scheme transforms the virtual MIMO channel into a set of parallel channels. Similarly, the achievable rate for the uniform input distribution can be expressed as\footnote{Dividing the rate by $1/2N$ serves two purposes; first, it accounts for the consideration of real signals (one dimension) instead of complex, and second, it aligns with the requirement of utilizing $N$ channel uses to achieve capacity.}
\begin{align}\tag{7}\label{mod6}
R_{u}^{\rm CI}(\mathbf{A}) = \sum_{i=1}^{N} \frac{1}{2N} \log_2\left( 1 + \frac{2\mathcal{E}_i^2}{\pi e \| \mathbf{b}_i\|^2} \right).
\end{align}

To gain more insight into the achievable rate expressions, we can compute the channel inverse $\mathbf{B}$ as
\begin{equation}\tag{8}\label{mod8}
\mathbf{B}=\mathbf{A}^{-1}=\left[\begin{array}{cccccc}
\frac{1}{\alpha h_1} & 0 & 0 & \cdots & 0 & 0 \\
\frac{\beta-1}{\alpha h_2} & \frac{1}{\alpha h_2} & 0 & \cdots & 0 & 0 \\
0 & 	\frac{\beta-1}{\alpha h_3} & \frac{1}{\alpha h_3} & \cdots & 0 & 0 \\
\vdots & \vdots & \vdots & \ddots & \vdots& \vdots \\
0 & 0 & 0 & \cdots & \frac{1}{\alpha h_{N-1}}& 0\\
0 & 0 & 0 & \cdots & \frac{\beta-1}{\alpha h_{N-1}}& \frac{1}{\alpha h_N}
\end{array}\right].
\end{equation}
Thus, we obtain
\begin{equation}\tag{9}\label{mod9}
\frac{1}{\|\mathbf{b}_i\|^{2}}=\begin{cases}
\alpha^2 h_1^2,& i =1,\\[2mm]
\frac{\alpha^2 h_{i}^2}{1+(\beta-1)^2},& 2\leq i\leq N.
\end{cases}
\end{equation}
Ultimately, by using the above result, we can respectively rewrite \eqref{mod5} and \eqref{mod6} as
\begin{equation}\tag{10}\label{mod10}
\begin{aligned}
R_{e}^{\rm CI}(\mathbf{A})=&\frac{1}{2N}\left(\log_2\left(1+\frac{e\alpha^2 h_1^2\mathcal{E}_1^2}{2\pi}\right) \right. \\[1mm]
& \left. \qquad \ + \sum_{i=2}^{N}\log_2\left(1+\frac{e\alpha^2 h_i^2\mathcal{E}_i^2}{2\pi((\beta-1)^2+1)}\right)\right),
\end{aligned}
\end{equation}
and
\begin{equation}\tag{11}\label{mod11}
\begin{aligned}
R_{u}^{\rm CI}(\mathbf{A})=&\frac{1}{2N}\left(\log_2\left(1+\frac{2\alpha^2 h_1^2\mathcal{E}_1^2}{\pi e}\right) \right. \\[1mm]
& \left. \qquad \ + \sum_{i=2}^{N}\log_2\left(1+\frac{2\alpha^2 h_i^2\mathcal{E}_i^2}{\pi e((\beta-1)^2+1)}\right)\right),
\end{aligned}
\end{equation}
where the parameters related to the temperature channel are explicitly shown. Finally, by averaging over the fading channel, the ergodic achievable rate using channel inversion is obtained as
\begin{equation}\tag{12}\label{mod 12}
R_{d,\mathrm{erg}}^{\mathrm{CI}}=\mathbb{E}_{\bf A}\left\{R_d^{\rm CI}(\mathbf{A})\right\},
\end{equation}
where $d \in \left\{ e, u\right\}$, and the expectation over $\bf A$ is numerically evaluated.

\subsection{Upper Bound on the Capacity}
In the context of MIMO communication, QR decomposition is often employed to compute an upper bound on the channel capacity where the channel matrix $\mathbf{A}$ is decomposed into the product of an orthogonal matrix $\mathbf{Q}$ and an upper triangular matrix $\mathbf{R}$. Additionally, given that our proposed temperature-based virtual MIMO channel closely resembles the intensity MIMO channel (as the temperature is always positive), an upper bound on the ergodic capacity can be expressed as \cite{CHA}
\begin{equation}\tag{13}\label{mod13}
C_{\mathrm{erg}}=\mathbb{E}_{\bf{A}}\left[ \frac{1}{2} \log_2 \left(\frac{\prod_{i=1}^N s_{i, i}}{\text{det}(\bf{S})}\right) +  \frac{1}{N}\sum_{i=1}^N \bar{r}\left(s_{i, i}^{-\frac{1}{2}},\mathcal{E}_i\right)\right],
\end{equation}
where $s_{i,i}$ is the $i$-th diagonal entry of $\mathbf{S}\triangleq \mathbf{R}^{-1} \mathbf{R}^{-T}$. We simplify the expression by recognizing that ${\rm{det}}(\bf{Q})=1$ due to $\mathbf{Q}$ being orthogonal, allowing us to directly define $\mathbf{S}$ as $\mathbf{S}\triangleq \left(\mathbf{A}^{T} \mathbf{A}\right)^{-1}$. In essence, by exploiting the properties of QR decomposition and by capitalizing on the structure of the temperature channel matrix $\mathbf{A}$, QR decomposition becomes unnecessary. Furthermore, a capacity upper bound has been derived in \cite[Theorem 7]{LA} under an average power constraint. Thus, $\bar{r}\left(\cdot,\cdot\right)$ in \eqref{mod13} is given by\vspace*{-2mm}
\begin{equation}\tag{14}\label{mod14}
\begin{aligned}
\bar{r}\left(s_{i, i}^{-\frac{1}{2}},\mathcal{E}_i\right)  \leq & \log_2 \left(\gamma e^{-\frac{\delta^2}{2 \sigma^2}}+\sqrt{2 \pi} \sigma Q\left(\frac{\delta}{\sigma}\right)\right)+\frac{1}{2} Q\left(\frac{\delta}{\sigma}\right) \\[2mm]
& +\frac{\delta e^{-\frac{\delta^2}{2 \sigma^2}}}{2 \sqrt{2 \pi} \sigma}+\frac{\delta^2}{2 \sigma^2}\left(1-Q\left(\frac{\delta+s_{i, i}^{-\frac{1}{2}}\mathcal{E}_i}{\sigma}\right)\right) \\[2mm]
& +\frac{1}{\gamma}\left(\delta+s_{i, i}^{-\frac{1}{2}}\mathcal{E}_i+\frac{\sigma e^{-\frac{\delta^2}{2 \sigma^2}}}{\sqrt{2 \pi}}\right) \!\! -\!\frac{1}{2} \log_2 2 \pi e \sigma^2,
\end{aligned}
\end{equation}
where $\gamma >0$ and $\delta \geq 0$ are free parameters. We restrict our attention to this bound since it has been shown to be fairly tight at high SNR regimes. Ideally, $\bar{r}\left(\cdot,\cdot\right)$ should be numerically minimized over allowed values of $\gamma$ and $\delta$. Nevertheless, a suboptimal choice for the parameters $\gamma$ and $\delta$ was found in \cite{LA} and is given by \vspace*{-2mm}
\begin{equation}\tag{15}\label{mod15}
\begin{aligned}
    \gamma=&\frac{1}{2}\left(\! \delta+s_{i, i}^{-\frac{1}{2}}\mathcal{E}_i+\frac{\sigma e^{-\frac{\delta^2}{2 \sigma^2}}}{\sqrt{2 \pi}} \! \right)\!+\!\frac{1}{2} \! \left[\left(\! \delta\!+\!s_{i, i}^{-\frac{1}{2}}\mathcal{E}_i\!+\!\frac{\sigma e^{-\frac{\delta^2}{2 \sigma^2}}}{\sqrt{2 \pi}} \! \right)^{\!2}\right. \\[2mm]
    &+\left. 4\left(\delta+s_{i, i}^{-\frac{1}{2}}\mathcal{E}_i+\frac{\sigma e^{-\frac{\delta^2}{2 \sigma^2}}}{\sqrt{2 \pi}} \right) \sqrt{2 \pi} \sigma e^{\frac{\delta^2}{2 \sigma^2}} Q\left(\frac{\delta}{\sigma}\right)\right]^{1/2},
\end{aligned}
\end{equation}
and
\begin{equation}\tag{16}\label{mod16}
    \delta = \sigma \log_2 \left( 1 + \frac{s_{i, i}^{-\frac{1}{2}}\mathcal{E}_i}{\sigma} \right).
\end{equation}

\subsection{Harvested Energy}
We analyze the performance in terms of the average harvested energy. We define the average harvested energy as the amount of energy harvested by the receiver per unit time. According to the piecewise linear harvesting model in \eqref{EH_mod}, the average harvested energy is expressed as
\begin{equation}\tag{17}\label{avg_harv}
	\bar{E} = \mathbb{E}\{ \mathcal{Q}(P_i)\} = \int\nolimits_{P_{\rm th}}^{P_{\rm sat}}\eta xf_X(x)\mathrm{d}x+\int\nolimits_{P_{\rm sat}}^{\infty}\eta P_{\rm sat}f_X(x)\mathrm{d}x,
\end{equation}where $f_X(x)$ denotes the probability density function (pdf) of the received power. More specifically, the harvested energy depends on the actual received power, which is the product of the exponentially distributed channel power gain with the exponentially/uniformly distributed input power, \textit{i.e.}, $P_i=h_i S_i$. Therefore, it is essential to derive the product distribution. The process involves computing the pdf by obtaining the cumulative distribution function of the received power, conditioned on the input distribution. Thus, the pdf of the received power for an exponential input distribution with mean $\mathcal{E}_i$ is given by
\begin{equation}\tag{18}\label{pdf_exp}
	f_X(x)=\int\nolimits_0^{\infty}\frac{1}{S_i \mathcal{E}_i} \exp\left(-\frac{S_i}{\mathcal{E}_i}\right)\exp\left(-\frac{x}{S_i}\right)\mathrm{d}S_i.
\end{equation}
By plugging \eqref{pdf_exp} into \eqref{avg_harv} and after a few algebraic manipulations, the average harvested energy for the exponential input distribution is given in closed form by
\begin{equation}\tag{19}
\begin{aligned}
    \bar{E}=\frac{2 \eta}{\sqrt{\mathcal{E}_i}} \!\!&\left[ P_{\rm th}^{3/2} K_1\left(2\sqrt{\frac{P_{\rm th}}{\mathcal{E}_i}}\right) \right. \\
    & \left. - \sqrt{P_{\rm sat}}\left(P_{\rm sat}\!-\!1\right)K_1\left(2\sqrt{\frac{P_{\rm sat}}{\mathcal{E}_i}}\right) \right. \\
    & \left. + \sqrt{\mathcal{E}_i}\!\left(\!P_{\rm th}K_2\! \left(\! 2\sqrt{\frac{P_{\rm th}}{\mathcal{E}_i}}\right)\!-\!P_{\rm sat}K_2\! \left(2 \sqrt{\frac{P_{\rm sat}}{\mathcal{E}_i}}\right)\!\right)\right].
\end{aligned}
\end{equation}Similarly, for a uniform input distribution, the pdf of the received power is
\begin{equation}\tag{20}
	f_X(x)=\int\nolimits_0^{2\mathcal{E}_i}\frac{1}{2\mathcal{E}_i S_i} \exp\left(-\frac{x}{S_i}\right)\mathrm{d}S_i,
\end{equation}
where the average energy is given in closed form by
\begin{equation}\tag{21}
\begin{aligned}
    \bar{E}=\eta P_{\rm sat} e^{-\frac{P_{\rm sat}}{2\mathcal{E}_i}}\!-\!\frac{\eta}{4 \mathcal{E}_i} \!\!&\left[ P_{\rm th}^{2} \Gamma\left(0,\frac{P_{\rm th}}{2\mathcal{E}_i}\right) + P_{\rm sat}^{2}\Gamma\left(0,\frac{P_{\rm sat}}{2\mathcal{E}_i}\right) \right. \\
    & \left. + 4\mathcal{E}_i^2 \!\left(\!\Gamma\left(2,\frac{P_{\rm sat}}{2\mathcal{E}_i}\right) \!-\! \Gamma\left(0,\frac{P_{\rm th}}{2\mathcal{E}_i}\right)\right)\right].
\end{aligned}
\end{equation}

\section{Numerical Results}
\begin{figure*}
\centering
\begin{subfigure}{.49\textwidth}
  \centering
	\includegraphics[width=0.80\textwidth]{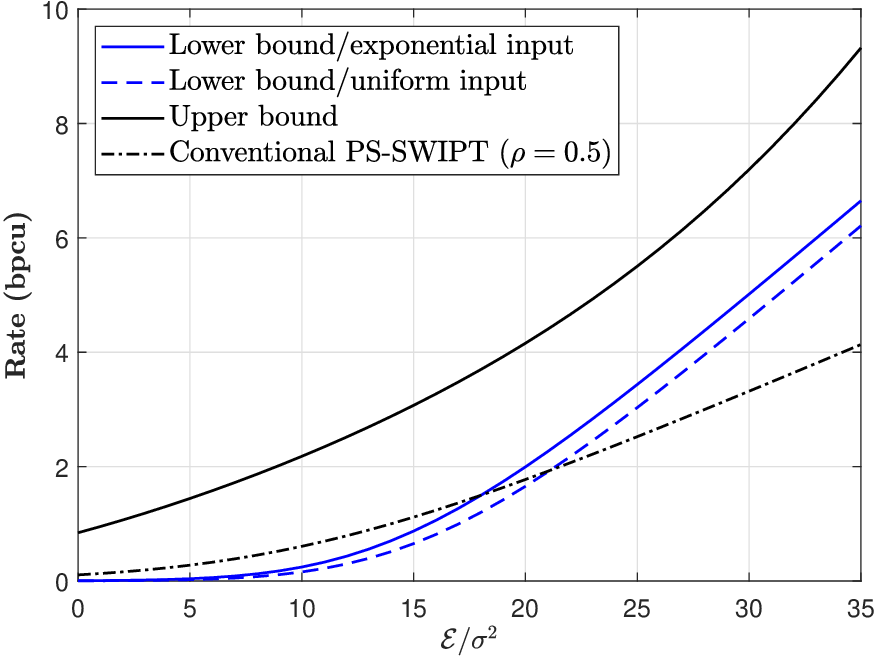}
 \vspace{-1mm}
	\caption{$N=4$.}\label{fig:Achiev_rate_N4}
\end{subfigure}%
\begin{subfigure}{.49\textwidth}
  \centering
	\includegraphics[width=0.80\textwidth]{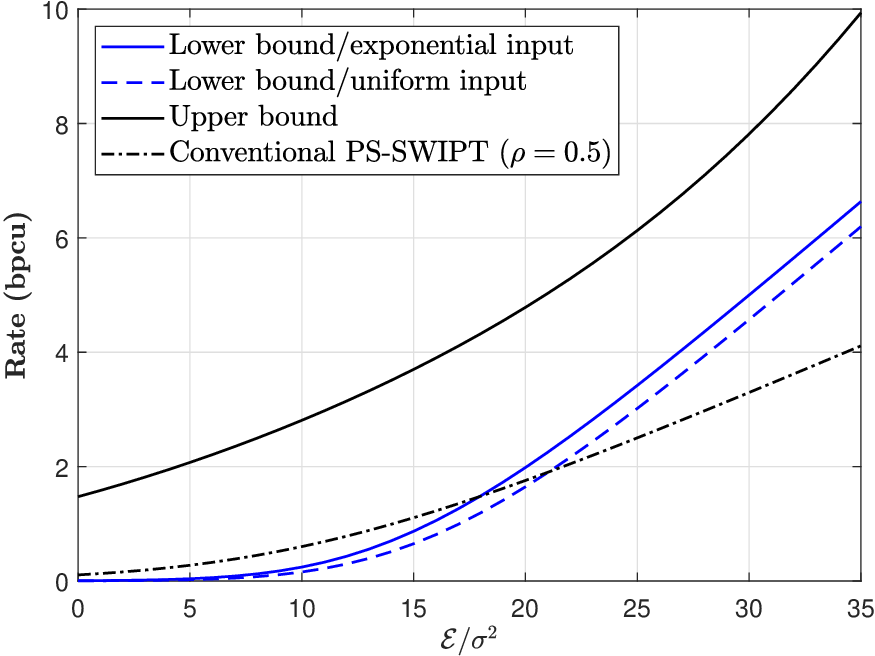}
 \vspace{-1mm}
	\caption{$N=6$.}\label{fig:Achiev_rate_N6}
\end{subfigure}
\vspace{-1mm}
\caption{Ergodic achievable rates and upper bound for the MIMO temperature channel under Rayleigh fading.}
\label{fig:Achiev_rate}
\vspace{-4mm}
\end{figure*}

\begin{figure}[t]
    \centering
    \includegraphics[width=0.40\textwidth]{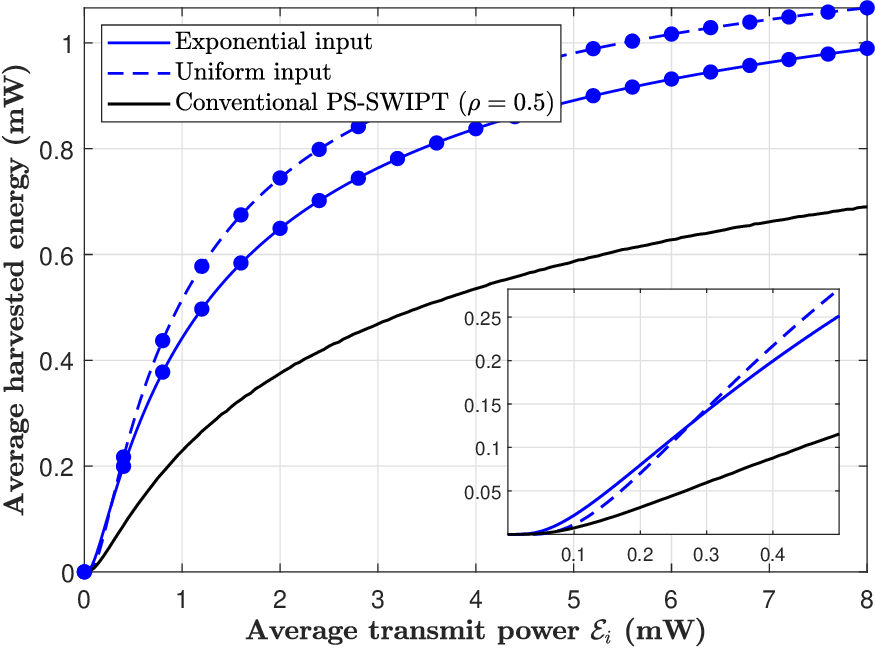}
    \caption{Average harvested energy performance. Lines and markers depict theoretical and simulation results, respectively.}
    \label{fig3}
    \vspace{-4mm}
\end{figure}

In this section, we provide numerical results to quantify the performance of the proposed temperature channel for SWIPT. Unless otherwise stated, we set $\mathcal{E}_i=\mathcal{E}, \forall i=1,\ldots, N$, and $\sigma^2 =1$. We assume $\alpha=0.1$, and $\beta=1-e^{-0.3}$ \cite{OZ2}. As for the EH process, we use $\eta=0.86$, $P_{\rm th}=0.5$ mW, and $P_{\rm sat}=1.5$ mW.

We first present the ergodic capacity upper bound $C_{\mathrm{erg}}$ in \eqref{mod13} along with the ergodic achievable rates $R_{e,\mathrm{erg}}^{\mathrm{CI}}$ and $R_{u,\mathrm{erg}}^{\mathrm{CI}}$ in Fig. \ref{fig:Achiev_rate}. We consider that the transmitter communicates with the receiver for two different channel use lengths, \textit{i.e.}, $N = \{4,6\}$. Hence, we obtain an $N \times N$ virtual MIMO temperature channel matrix. Furthermore, the achievable rates are plotted versus the signal-to-noise ratio (SNR). We observe from the results that the rates achieved by an exponential input distribution are higher than those achieved by considering a uniform input. This outcome is expected since the distribution that maximizes mutual information for nonnegative inputs is the exponential input distribution \cite[Sec. IV-A]{CHA1}; thus, opting for a uniform distribution is considered suboptimal in this context. Besides, we notice from both figures that the achievable rate does not increase with increasing $N$, \textit{i.e.}, from 4 to 6 channel uses. This behavior is attributed to the requirement of $N$ consecutive channel uses to achieve capacity. More specifically, the achievable rate expressions in \eqref{mod10} and \eqref{mod11} indicate a division by $N$ within the expressions, which implies a consistent ergodic rate per channel use. In addition, we consider a conventional power splitting (PS) SWIPT scheme with a PS ratio $\rho=0.5$ to ensure equal resource allocation between information decoding and EH \cite{ZH}. For this case, we assume a flat Rayleigh fading channel and a Gaussian input distribution, which is optimal for the communication transmission. Our proposed scheme exhibits superior performance compared to the PS-SWIPT technique, particularly at moderate to high SNRs.

Fig. \ref{fig3} depicts the average harvested energy performance versus the average transmit power for an exponential and uniform input distribution for the nonlinear EH model in $\eqref{EH_mod}$. The main observation is that the exponential input distribution exhibits a higher EH performance than the uniform input distribution for low input powers, while the reverse is observed for high input power regimes. This is because the exponential distribution yields higher average harvested energy at lower average input power due to its long tail allowing for occasional high power inputs that exceed the saturation point. On the other hand, the uniform distribution outperforms the exponential distribution at higher average input power, since its power inputs fall within the optimal EH range. We conclude that the selection of the input distribution affects the system performance, emphasizing that, depending on the operating input power regime, enhancements in rate may come at the expense of EH, and vice versa. Concerning the comparison with the PS-SWIPT, we remark that our proposed scheme exhibits superior EH performance since the conventional technique suffers from a fundamental tradeoff between information and energy. Finally, the simulation results (markers) are in line with our theoretical analysis (lines).

\section{Conclusion}
In this work, we introduced a novel communication paradigm to SWIPT systems by harnessing the thermal effects induced by electromagnetic signals. Unlike traditional methods, the temperature-based SWIPT system avoids resource division at the receiver. By exploiting the temporal dimension of the temperature channel, we constructed an equivalent virtual MIMO channel. Our derived achievable rate and EH analysis highlighted the impact of input distribution on performance, particularly in scenarios with moderate to high input power, offering valuable design insights for temperature-based SWIPT systems. The new SWIPT model is promising for future applications such as EH nano-sensor networks.


\begin{thebibliography}{5}
\bibitem{OZ} O. Ozel, K. Tutuncuoglu, S. Ulukus, and A. Yener, ``Fundamental limits of energy harvesting communications,'' {\it IEEE Commun. Mag.}, vol. 53, no. 4, pp. 126-1329, Apr., 2015.

\bibitem{LU} X. Lu, P. Wang, D. Niyato, D. I. Kim, and Z. Han, ``Wireless networks with RF energy harvesting: A contemporary survey,'' {\it IEEE Commun. Surveys Tuts.}, vol. 17, no. 2, pp. 757-789, 2nd Quart., 2015. 

\bibitem{KRI} I. Krikidis, S. Timotheou, S. Nikolaou, G. Zheng, D. W. K. N, and R. Schober, ``Simultaneous wireless information and power transfer in modern communication systems,'' {\it IEEE Commun. Mag.}, vol. 52, no. 11, pp. 104-110, Nov., 2014.

\bibitem{PE} T. D. P. Perera, D. N. K. Jayakody, S. K. Sharma, S. Chatzinotas, and J. Li, ``Simultaneous wireless information and power transfer (SWIPT): Recent advances and future challenges,'' {\it IEEE Commun. Surveys Tuts.}, vol. 20, no. 1, pp. 264-302, 1st Quart., 2018.

\bibitem{WEI} Z. Wei, X. Yu, D. W. K. Ng, and R. Schober, ``Resource allocation for simultaneous wireless information and power transfer systems: A tutorial overview,'' {\it Proc. IEEE}, vol. 110, no. 1, pp. 127-149, Jan. 2022.

\bibitem{ZA} S. Zander, P. Branch, and G. Armitage, ``Capacity of temperature-based covert channels,'' {\it IEEE Commun. Lett.}, vol. 15, no. 1, pp. 82-84, Jan. 2011.

\bibitem{OZ2} O. Ozel, S. Ulukus, and P. Grover, ``Energy harvesting transmitters that heat up: Throughput maximization under temperature constraints,'' {\it Trans. Wireless Commun.}, vol. 15, no. 8, pp. 5440-5452, Aug. 2016.

\bibitem{TO} R. G. Gebremedhin, and T. L. Marzetta, ``Thermal conduction as a wireless communication channel,'' in {\it Proc. IEEE Glob. Commun. Conf}, pp. 1085-1090, Rio De Janeiro, Brazil, Dec. 2022.

\bibitem{CHA1} A. Chaaban, Z. Rezki, and M.-S. Alouini, ``On the capacity of intensity-modulation direct-detection Gaussian optical wireless communication channels: A tutorial,'' {\it IEEE Commun. Surveys Tuts.}, vol. 24, no. 1, pp. 455-491, 1st Quart. 2009.

\bibitem{CHA} A. Chaaban, Z. Rezki, and M.-S. Alouini, ``MIMO intensity-modulation channels: Capacity bounds and high SNR characterization,'' {\it in Proc. IEEE Int. Conf. Commun. (ICC)}, Paris, France, May 2017, pp. 1-6.

\bibitem{LA} A. Lapidoth, S. M. Moser, and M. A. Wigger, ``On the capacity of free-space optical intensity channels,'' {\it IEEE Trans. Inf. Theory}, vol. 55, no. 10, pp. 4449-4461, Oct. 2009.

\bibitem{ZH} X. Zhou, R. Zhang, and C. K. Ho, ``Wireless information and power transfer: Architecture design and rate-energy tradeoff,'' {\it IEEE Trans. Commun.}, vol. 61, no. 11, pp. 4754-4767, Nov. 2013.

\end{thebibliography}
\end{document}